\documentclass[english,pra,showpacs,showkeys,tightenlines,secnumarabic,11pt]{revtex4}
\usepackage[T1]{fontenc}
\usepackage[latin1]{inputenc}
\usepackage{amsmath}
\usepackage{graphicx}
\usepackage{amssymb}
\usepackage{epsfig}
\usepackage{graphics}
\usepackage[mathscr]{euscript}
\usepackage{psfrag}
\usepackage{pstricks}
\usepackage{pst-node}
\newcommand{\noun}[1]{\textsc{#1}}
\usepackage{babel}
\begin{document}
\title{\bf  Cronin effect and energy conservation constraints  in high
energy proton-nucleus collisions}
%  \vspace{1.5cm}
\author{Enrico Cattaruzza}
\email{ecattar@ts.infn.it}
\author{Daniele Treleani}
\email{daniel@ts.infn.it}
\affiliation{Dipartimento di Fisica Teorica dell'Universit\`a di Trieste and INFN, Sezione di Trieste,\\Strada Costiera 11, Miramare-Grignano, I-34014 Trieste,Italy.}
\date{\today}
\begin{abstract}
We estimate the Cronin effect in $pA$ collisions at the CERN LHC and at RHIC, using a
Glauber-Eikonal model of initial state multiparton interactions. For a correct
determination of the initial parton flux, we upgrade the model cross section, taking
carefully into account all kinematical constraints of each multi-parton interaction
process. As compared with previous results, derived with approximate kinematics, we
obtain a softer spectrum of produced partons, while improving the agreement of the model
with the recent measurements of $\pi^0$ production in $d+Au$ collisions at $\sqrt s=200$
AGeV.
\end{abstract}
\maketitle
\section{Introduction}

Hadron-nucleus interactions represent an intermediate regime between hadron-hadron and
nucleus-nucleus collisions, where different ideas on the role played by the complexity of
the target on large $p_{t}$ dynamics may be tested~\cite{Accardi:2003be}. Of particular
interest is the study of the Cronin effect~\cite{Antreasyan:cw,Straub:xd}, namely the
deformation of the transverse spectrum of particles, or jets, produced in hadron-nucleus
collisions in the projectile fragmentation region. Given the large scale of momentum
transfer in the process, the problem is within reach of a perturbative approach, while
the effects of the complex structure of the target may be controlled by changing energy
and atomic mass number.

The effect is originated by the high density of the nuclear target, which induces the
projectile's partons interacting simultaneously with several partons of the target. There
is not however a unique explanation of the underlying dynamical
mechanism~\cite{K.J.Eskola,Wang:1998ww,Wang:2001cy,Zhang:2001ce,Kopeliovich:2002yh} and
different approaches can only be discriminated by measuring the large $p_{t}$ spectra in
hadron nucleus collisions at the energies of the future colliders, where various models
give different predictions. In our opinion, the simplest approach is to adopt the Glauber
prescription of factorization of the overall many-parton $S$ matrix, as a product of
elementary partonic $S$ matrices~\cite{Braun:2002kg}. Since production at the parton
level represents a higher order correction in the coupling constant, the natural zeroth
order approximation is obtained when evaluating each elementary parton interaction at the
lowest order in $\alpha_{S}$, which leads considering binary processes only.

Encouragingly a recent calculation, based on this simplest scheme~\cite{Accardi:2003jh},
compares rather reasonably with the latest experimental measurements of the Cronin effect
in large $p_{t}$ $\pi^0$ production in $d+Au$ collisions, at a center of mass energy of
$200$ AGeV \cite{Adler:2003ii}, \cite{Arsene:2003yk}, \cite{Back:2003ns},
\cite{Adams:2003im}, which represents a convenient channel to study the effect (as,
differently from charged particle production, there is no contamination from protons,
whose production mechanism poses additional problems).

It's remarkable that in such an approach, when adopting the usual approximations of the
Glauber picture of high energy hadron-nucleus collisions, the whole effect of the nuclear
target medium may be taken into account exactly, in the evaluation of the inclusive
transverse spectrum~\cite{Braun:2002kg}. After summing all possible multi-parton
interactions of the projectile parton one in fact obtains

\begin{equation}
\frac{d\sigma}{d^{2}bdxd^{2}p_{t}}=\frac{1}{(2\pi)^{2}}\int d^{2}re^{ip_{t}r}G(x)\,
S_{hard}^{A}(\bar{r},\,\bar{b},\, p_{0})\label{summed}\end{equation} where

\[
S_{hard}^{A}(\bar{r},\,\bar{b},\,
p_{0})=\left[e^{T_{A}(b)[F_{A}(x,r)-F_{A}(x,0)]}-e^{-T_{A}(b)F_{A}(x,0)}\right]\]

\noindent and $T_{A}(b)$ is the usual nuclear thickness, as a function of the
hadron-nucleus impact parameter $b$, $G(x)$ the parton number density of the projectile,
as a function of the fractional momentum $x$, $p_{t}$ the transverse momentum of the
final observed parton and

\begin{equation}
F_{A}(x,b)=\int\frac{d^{2}p_{t}}{(2\pi)^{2}}dx'G_{A}(x')\frac{d\hat{\sigma}}{d^{2}p_{t}}e^{ip_{t}b}\label{FA}\end{equation}

In Eq.(\ref{FA}) $G_{A}(x')$ is the parton number density of the target nucleus, divided
by the atomic mass number $A$, as a function of the fractional momentum $x'$, and
$d\hat{\sigma}/d^{2}p_{t}$ the elementary parton-parton scattering cross section, which
includes also the kinematical limits constraints. For sake of simplicity flavor
dependence is not written explicitly. The quantity $S_{hard}^{A}(\bar{r},\,\bar{b},\,
p_{0})\,$can be interpreted in terms of dipole-nucleus hard cross section. The dipole
originates from the square of the scattering amplitude, expressed as a function of the
Fourier variable $\bar{r}\,,$ corresponding to the transverse size of the dipole
\cite{Accardi Colour}.

\noindent By expanding Eq.(\ref{summed}) as a power series in $\hat{\sigma}$, one obtains
all different multiple scatterings of the projectile parton. As a consequence of the
Glauber prescription of factorization of the overall many-parton $S$ matrix, each term in
the expansion is represented by an incoherent convolution of binary partonic
collisions~\cite{Braun:2002kg,Accardi:2001ih}. Analogously to the case of the Glauber
approach to hadron-nucleus processes, the $S$-matrix factorization prescription does not
however imply any space-time ordering between the different elementary collisions. It
only implies that a connected $n$-body interaction process is well approximated by a
product of two-body interactions, which corresponds to the dominance of the pole
contribution, in a dispersive representation of the projectile-exchanged gluon amplitude
\cite{Braun:pk}.

All expressions are divergent for small values of the lower cutoff in $p_t$, which sets
the regime of hard interactions. Nevertheless, as unitarity is explicitly implemented,
each term in the series expansion of Eq.(\ref{summed}) contains subtractions, which lower
the degree of the infrared singularity, from an inverse power to a power of a logarithm
of the cutoff \cite{Accardi:2001ih}. It's also worth mentioning that, in this model, the
average energy, carried by the partons scattered to the final state is bound to be
smaller than the overall initial state energy \cite{Accardi:2000ry}, whereas such a
quantity is divergent as an inverse power of the cutoff, if evaluated in impulse
approximation.

The approach to hard processes in hadron-nucleus collisions, based on the factorization
of the multi-parton $S$ matrix, is hence unitary and satisfies various non-trivial
consistency requirements, including the AGK cutting rules \cite{AGK}. As
$p_{t}\rightarrow0$ the importance of unitarity corrections grows, producing a
suppression of the integrated parton yield and a random walk of partons to higher
$p_{t}\,,$ recovering in this way the local isotropy in transverse space of the black
disk limit, which is, on the contrary, maximally broken by the lowest order impulse
approximation term \cite{Accardi:2000ry}.

The model hence contains geometrical shadowing at low $p_{t}$ and Cronin enhancement of
the transverse spectrum at moderate $p_{t}$. Some non secondary features of the
interaction are not accounted however, as one expects genuine dynamical shadowing at low
$x$, due to non-linear gluon interactions \cite{Accardi:2003be}. In spite of some
interesting efforts in this direction \cite{Kovchegov:1998bi}, incorporating non-linear
gluon evolution in the present formalism is however a non trivial problem, which we will
not try to address in the present note.

In fact we aim to point out that also some of the simplifying assumptions, adopted in the
derivation of Eq.(\ref{summed}), require a closer inspection. More specifically, to
obtain Eq.(\ref{summed}), one needs to decouple longitudinal and transverse degrees of
freedom, while conserving the longitudinal momentum components in the interaction.
Kinematical constraints are hence implemented only approximately, in such a way that
final state partons turn out to be more energetic than the initial ones, the effect being
emphasized when the number of re-scatterings grows. The final $p_{t}$ spectrum gets hence
shifted towards larger transverse momenta which, given the steepness of parton
distributions at small $x$, might modify appreciably the evaluation of the Cronin effect.

The purpose of the present note is to investigate this specific aspect of the model more
closely. We hence compare the spectrum, obtained by implementing exactly all kinematical
constraints, with the summed expression in Eq.(\ref{summed}), both at the LHC regime and
in the kinematical domain of the recent measurements of the Cronin effect in
$d+Au\to\pi^0X$ at RHIC.

\section{Transverse Spectrum}

\noindent The expression in Eq.(\ref{summed}) represents the sum of all possible
collisions of the observed parton with the different scattering centers of the target
nucleus. When the kinematical constraints are implemented exactly, the multi-scattering
series cannot be re-summed any more in a closed analytic expression and each different
term needs to be evaluated separately. Most of the spectrum is nevertheless well reproduced
by the the first three terms of the expansion \cite{Accardi:2001ih}. Hence we approximate
the cross section by:

\begin{eqnarray}
\frac{d\sigma}{d^{2}p_{t}dy\, d^{2}b} & = & \sum_{i}\frac{d\sigma_{i}}{d^{2}p_{t}dy\,
d^{2}b}\label{eq:single+double+triple}\end{eqnarray} where $p_{t}$ is the transverse
momentum of the observed parton, $y$ its rapidity, $b$ the impact parameter of the
collision and the sum runs over the different species of projectile partons, while

\begin{equation}
\frac{d\sigma_{i}}{d^{2}p_{t}dy\, d^{2}b}=\frac{d\sigma_{i}^{(1)}}{d^{2}p_{t}dy\,
d^{2}b}+\frac{d\sigma_{i}^{(2)}}{d^{2}p_{t}dy\,
d^{2}b}+\frac{d\sigma_{i}^{(3)}}{d^{2}p_{t}dy\,
d^{2}b}\label{eq:triple-expansion}\end{equation} where

\begin{eqnarray*}
\frac{d\sigma_{i}^{(1)}}{d^{2}p_{t}dy\, d^{2}b} & = & T_{A}(b)\,\sum_{j}\,\frac{1}{1+\delta_{ij}}\int\; d^{2}q_{1}\, dx'\,\delta^{(2)}(\bar{q}_{1}-\bar{p}_{t})\,\hat{\sigma_{ij}}^{(1)}(y,x';q_{1})\\
 & \times & \, x\,\left[f_{\frac{i}{p}}(x,\, Q_{fct})\, f_{\frac{j}{A}}(x',\, Q_{fct})+f_{\frac{i}{p}}(x',\, Q_{fct})\, f_{\frac{j}{A}}(x,\, Q_{fct})\right]\end{eqnarray*}
is the single scattering term (the projectile parton interacts with a single parton of
the target and vice-versa),

\begin{eqnarray*}
\frac{d\sigma_{i}^{(2)}}{d^{2}p_{t}dy\, d^{2}b} & = & \frac{1}{2!}\: T_{A}(b)^{2}\,\sum_{j\,,k}\,\frac{1}{1+\delta_{ij}}\,\frac{1}{1+\delta_{ik}}\,\int\, d^{2}q_{1}\, d^{2}q_{2}\, dx'\, dx''\\
 & \times & \left[\delta^{(2)}(\bar{q}_{1}+\bar{q}_{2}-\bar{p}_{t})-\delta^{(2)}(\bar{q}_{1}-\bar{p}_{t})-\delta^{(2)}(\bar{q}_{2}-\bar{p}_{t})\right]\\
 & \times & \, x\, f_{\frac{i}{p}}(x,\, Q_{fct})\, f_{\frac{j}{A}}(x',\, Q_{fct})\, f_{\frac{k}{A}}(x'',\, Q_{fct})\,\hat{\sigma}_{ijk}^{(2)}(y,x',x'';\bar{q}_{1},\bar{q}_{2})\end{eqnarray*}
is the rescattering term (the projectile parton interacts with two different partons of
the target), and

\begin{eqnarray*}
\frac{d\sigma_{i}^{(3)}}{d^{2}p_{t}dy\, d^{2}b} & = & \frac{1}{3!}\: T_{A}(b)^{3}\,\sum_{j\,,k,\, l}\,\frac{1}{1+\delta_{ij}}\,\frac{1}{1+\delta_{ik}}\,\frac{1}{1+\delta_{il}}\,\int\, d^{2}q_{1}\, d^{2}q_{2}\, d^{2}q_{3}\\
 & \times & \, dx'\, dx''\, dx'''\,\hat{\sigma}_{ijkl}^{(3)}(y,x',x'',x''';\bar{q}_{1},\,\bar{q}_{2},\,\bar{q}_{3})\\
 & \times & \left[\delta^{(2)}(\bar{q}_{1}+\bar{q}_{2}+\bar{q}_{3}-\bar{p}_{t})-\delta^{(2)}(\bar{q}_{1}+\bar{q}_{2}-\bar{p}_{t})-\delta^{(2)}(\bar{q}_{2}+\bar{q}_{3}-\bar{p}_{t})\right.\\
 && \left.-\delta^{(2)}(\bar{q}_{3}+\bar{q}_{1}-\bar{p}_{t})+\delta^{(2)}(\bar{q}_{1}-\bar{p}_{t})+\delta^{(2)}(\bar{q}_{2}-\bar{p}_{t})+\delta^{(2)}(\bar{q}_{3}-\bar{p}_{t})\right]\\
 & \times & x\, f_{\frac{i}{p}}(x,\, Q_{fct})\, f_{\frac{j}{A}}(x',\, Q_{fct})\, f_{\frac{k}{A}}(x'',\, Q_{fct})\, f_{\frac{l}{A}}(x''',\, Q_{fct}).\end{eqnarray*}
is the double rescattering term (the projectile parton interacts with three different
partons of the target).

\noindent The subtraction terms, in the rescattering and in the double rescattering
terms, are a direct consequence of the implementation of unitarity in the process. The
sums run over the different species of target partons. The quantities
$f_{\frac{j}{A}}(x',\, Q_{fct})$ represent the nuclear parton distributions, while the
multiparton interaction cross sections are given by:

\begin{eqnarray*}
\hat{\sigma_{ij}}^{(1)}(y,x';q_{1}) & =k_{factor}\, &
\frac{d\hat{\sigma}_{ij}(\hat{s}_{1},\,\hat{t}_{1},\,\hat{u}_{1})}{d\hat{t}_{1}}\,\Theta(\Phi^{(1)})\end{eqnarray*}

\begin{eqnarray*}
\hat{\sigma}_{ijk}^{(2)}(y,x',x'';\bar{q}_{1},\bar{q}_{2}) & = &
(k_{factor})^{2}\,\frac{d\hat{\sigma}_{ik}(\hat{s}_{2},\,\hat{t}_{2},\,\hat{u}_{2})}{d\hat{t}_{2}}\,\frac{d\hat{\sigma}_{ij}(\hat{s}_{1},\,\hat{t}_{1},\,\hat{u}_{1})}{d\hat{t}_{1}}\,\Theta(\Phi^{(2)})\,\Theta(\Phi^{(1)})\end{eqnarray*}

\begin{eqnarray*}
\hat{\sigma}_{ijkl}^{(3)}(y,x',x'',x''';\bar{q}_{1},\,\bar{q}_{2},\,\bar{q}_{3}) & = & (
k_{factor})^{3}\,\frac{d\hat{\sigma}_{il}(\hat{s}_{3},\,\hat{t}_{3},\,\hat{u}_{3})}
{d\hat{t}_{3}}\,\frac{d\hat{\sigma}_{ik}(\hat{s}_{2},\,\hat{t}_{2},\,\hat{u}_{2})}
{d\hat{t}_{2}}\\
 & \times & \frac{d\hat{\sigma}_{ij}(\hat{s}_{1},\,\hat{t}_{1},\,\hat{u}_{1})}{d\hat{t}_{1}}\
\Theta(\Phi^{(3)})\,\Theta(\Phi^{(2)})\,\Theta(\Phi^{(1)})\,,\end{eqnarray*} where the
parton$_i$-parton$_j$ differential cross sections $d\hat{\sigma}_{ij}/d\hat{t}$ are
evaluated at the lowest order in pQCD by making use of exact kinematics. The kinematical
constraints are implemented through the quantities $\Phi^{(i)}\,$. The expressions of
$\Phi^{(i)}\,$ and of the invariants $\hat{s}_{i}\,,\hat{t}_{i},\,\hat{u}_{i}$ are
derived in great detail in the appendix.

The infrared singularity is regularized by introducing in the propagators the mass
parameter $p_{0}\,,$ which is a free parameter in the model. Higher order-effects in the
elementary interaction are accounted by multiplying the lowest order expressions in
$\alpha_s$ by the factor $k_{fact}$.

When evaluating the $\pi^0$ spectrum, we let partons fragment independently, neglecting
the transverse momentum component generated by the fragmentation process. The relations
between hadronic and partonic variables are hence the following \cite{K.J.Eskola2002}:

\begin{eqnarray*}
m_{h}\cosh y_{h}=z\, p_{t}\,\cosh y &  & m_{h}\sinh y_{h}=q_{t}\,\sinh y\:\end{eqnarray*}
where $z=\frac{E_{h}}{E}\,$, with $E$ the energy of the parton, $y$ its rapidity, $E_{h}$
the energy of the hadron $h$, $m$ its mass, $y_h$ its rapidity, $q_{t}$ its transverse
momentum and $m_{h}$ its transverse mass. The inclusive $\pi^0$ spectrum is hence given
by:\[ \frac{d\sigma^{frag}}{d^{2}q_{t}dy_{h}\,
d^{2}b}=\sum_{i}\,\frac{d\sigma_{i}^{frag}}{d^{2}q_{t}dy_{h}\, d^{2}b}\] with

\begin{eqnarray*}
\frac{d\sigma_{i}^{frag}}{d^{2}q_{t}dy_{h}\, d^{2}b} & = &
J(m_{h},y_{h})\,\int\,\frac{dz}{z^{2}}\, D_{i\rightarrow
h}(z,Q_{F}^{2})\,\left.\frac{d\sigma_{i}}{d^{2}p_{t}dy\, d^{2}b}\right|_{p_{t},\,
y}\:,\end{eqnarray*} where \begin{eqnarray*}
p_{t}=\frac{q_{t}}{z}\, J(m_{h},y_{h})\qquad & y=\arcsin\left(\frac{m_{h}\,\sinh y}{q_{t}}\right)\\
J(m_{h},y_{h})= &
\left(1-\frac{m^{2}}{m_{h}^{2}\,\cosh^{2}y_{h}}\right)^{-\frac{1}{2}}\end{eqnarray*} and
$D_{i\rightarrow h}(z,Q_{F}^{2})$ the fragmentation functions, depending on the
fragmentation scale $Q_F$. The integration region for the hadron energy fraction $z$ is
set by the request $p_{t}\ge p_{0}$, with $p_{0}$ the infrared regulator:

\begin{eqnarray*}
\frac{2\: m_{h}\:\cosh y_{h}}{\sqrt{s}} & \le z\le & \min\left[1,\frac{q_{t}}{p_{0}}\,
J(m_{h},y_{h})\right]\,.\end{eqnarray*}
%%%%%%%%%%%%%%%%%%%%%%%%%%%%%%%%%%%%%%%%
\subsection{Numerical Results}
In our calculations we use the isospin averaged nuclear parton distributions,
$f_{\frac{i}{A}}=Z\, f_{\frac{i}{p}}+(A-Z)\, f_{\frac{i}{n}}\,$, where for the proton and
the neutron distributions, $f_{\frac{i}{p}}$ and $f_{\frac{i}{n}}$, we take the CTEQ5
parametrization at the leading order \cite{CTEQ5}, while for the nuclear thickness
function $T_A(b)$, which is normalized to one, we use the Wood-Saxon parametrization.

To study the effect at the LHC we consider the case of production of minijets in a
forward calorimeter ($\eta\in[2.4,\,4]$) at two different center of mass energies in
the hadron-nucleon c.m. system, $\sqrt{s}=5.5$ ATeV and $\sqrt{s}=8.8$ ATeV, which might
be compatible with the acceptances of ATLAS and CMS \cite{CMS}. Given the small
values of $x$ in this kinematical range, we use effective parton distribution functions,
with the gluon-gluon cross-section as a dynamical input. We set the factorization and
renormalization scale $Q_{fct},\, Q_{rn}\,$equal to the regularized transverse mass
$m_{t}=\sqrt{p_{0}^{2}+p_{t}^{2}}$ and we evaluate the transverse spectrum of jets, in
proton-lead collisions, by the expression:

\begin{eqnarray}
\frac{dW_{pA}}{d^{2}p_{t}} & = & \frac{1}{y_{max}-y_{min}}\,\int_{y\in[y_{min},\,
y_{max}]}\, dx\, d^{2}b\,\frac{d\sigma}{d^{2}p_{t}dy\,
d^{2}b};\label{eq:Rapidity-integrated-spectrum}\end{eqnarray} where the integrand is
given by Eq.(\ref{eq:single+double+triple}). As for the choice of the infrared regulator
we set $p_0=2\,GeV$    and we use $k_{factor}=2$. Our results are plotted in Fig.1, where
we compare the spectrum (\ref{eq:Rapidity-integrated-spectrum}) obtained by the exact
implementation of energy conservation in the multiple interactions (solid line), with the
approximate kinematics results given by the first three terms of the expansion of
Eq.(\ref{summed}). As an effect of the exact implementation of kinematics the spectrum of
the outgoing particle is shifted towards lower transverse momenta, resulting in an
appreciable energy loss effect: at $p_t\sim15\, GeV$ the energy-loss correction to the
spectrum is of the order of 40-50\%, at higher transverse momenta, $p_t\sim30\, GeV$,
it's still about 30-38\%. The triple scattering approximation, used to evaluate the
spectrum, breaks down at $p_t\le9\, GeV$ at $\sqrt{s}=8.8\,TeV$: by increasing the center
of mass energy the density of target partons grows rapidly and the contribution of higher
order rescatterings cannot be neglected any more at  $p_t\le9\, GeV$ (right panel of
Fig.1).
 As the energy is lowered to $\sqrt{s}=5.5\,TeV$, the
region of numerical instability is shifted to the region $p_{t}\le3\, GeV\,$(left panel
of Fig.1).

The softening of the spectrum has important effects on the Cronin ratio, defined as:
\[
R(p_{t})=\frac{dW_{pA}/d^{2}p_{t}}{dW_{pA}^{(1)}/d^{2}p_{t}}.\] where $dW_{pA}^{(1)}/
d^{2}p_{t}$ is the cross section in impulse approximation. The Cronin ratio,
corresponding to the spectra in fig.1, is shown in fig.2, where the dashed and continuous
lines correspond to approximate and exact kinematics respectively. The two panels of
fig.2 refer to the two values considered for the center of mass energy; at higher energy
multiple interactions produce a larger value of the Cronin ratio. The overall effect of
the implementation of energy conservation on $R$ is to decrease the size of the Cronin
effect and to shift the curve towards smaller transverse momenta.
%%%%%%%%

At lower energy, $\sqrt{s}=200\,$ AGeV, we follow \cite{Accardi:2003jh} in the evaluation
of the cross section of $d+Au\rightarrow\pi^{0}X$, using
$Q_{fct}=Q_{rn}=\frac{m_{t}}{2}\,$, and the values $p_{0}=1.0\,$ GeV and $k_{fact}=1.04$.
These choices, besides minimizes the effects of higher order corrections in $\alpha_S$,
allow one reproducing the inclusive cross section of $\pi^0$ production in $pp$
collisions at the same c.m. energy, without any need of smearing the cross section with
an intrinsic parton transverse momentum. At $\sqrt{s}=200\,$ AGeV we take explicitly into
account the differences between partons, both in the distributions and in the evaluation
of the partonic cross sections. The resulting Cronin ratio in $d+Au$ collisions is hence
a parameter free prediction of the our model. By using the leading order \emph{K-K-P}
fragmentation functions \cite{KKP} at the fragmentation scale $Q_{F}=\frac{m_{t}}{2}$, we
evaluate:
\[
R_{dAu\rightarrow\pi^{0}X}=\frac{d\sigma_{dAu\rightarrow\pi^{0}X}^{frag}}{d^{2}q_{t}\,
dy\, d^{2}b}\left/\frac{d\sigma_{dAu\rightarrow\pi^{0}X}^{frag\,(1)}}{d^{2}q_{t}\, dy\,
d^{2}b}\right.\] at $y=0$ and $b=b_{dAu}=5.7\, fm,\,$ which is the estimated average
impact parameter of the experiment \cite{bimpact}. In fig.3 we compare our result
(continuous line) with the experimental data of the PHENIX collaboration
\cite{Adler:2003ii} and with the result obtained by using the approximate kinematics of
Eq.(\ref{summed}), dashed line. Given the small rapidity values of the observed $\pi^0$,
in this case the corrections induced by exact kinematics are particularly important in
the region of smaller transverse momenta. The dependence of the effect on the impact
parameter is shown in fig.4, where our calculation (continuous line) is compared with
preliminary data presented by PHENIX at the DNP Fall Meeting, held in Tucson last October
2003 \cite{DNP}, and with the the standard Glauber-Eikonal calculation (dashed line). The
fractions refer to the BBC centrality selection. The corresponding estimated impact
parameter values are: 0-20\% => b=3.5 fm 20-40\% $\to$ b=4.5 fm 40-60\% $\to$ b=5.5 fm
60-88\% $\to$ b=6.5 fm. As an effect of implementing exactly kinematical constraints, one
obtains a reduction of the Cronin curve in the region of smaller $p_t$ values, resulting
with an improved agreement of the model with the experimental indication. Notice that
besides the deformation of the spectrum leading to geometrical shadowing, by implementing
energy conservation one obtains also the quenching the spectrum, as part of the
projectile's energy is transferred to the target.

The expectation for $\pi^0$ production in $d+Au$ at $\eta=3.2$ is shown in fig.5, where
we compare our result (continuous line) with the standard Glauber-Eikonal model
calculation (dashed line). Given the larger average number of rescatterings at larger
rapidity values, the quenching of the spectrum, due to the energy lost by the projectile
in the multiple collisions, is now sizably increased (notice that the vertical scale is
now different with respect to the previous figures 3 and 4). One in fact obtains values
of $R_{dAu\rightarrow\pi^{0}X}$ as small as $.6$ at $p_t=1$ GeV. The expectation is
nevertheless that the Cronin curve should exceed one for $p_t\ge2$ GeV, which is not the
case for charged particle production. In the figure we plot preliminary results for
charged particle production at $\eta=3.2$, presented by BRAHMS at the Quark Matter
Conference\cite{Brahms}. In the forward region the quenching of the spectrum might hence
be a sizably larger effect, not accountable by multiple scatterings.
%%%%%%%%%%%%%%%%%%%%%%%%%%%%%%%%%%%%%%%%%%%%%
\subsection{Conclusions}
In very high energy proton-nucleus collisions the interaction probability is very large,
also when considering processes with momentum transfer well above the lower limit for a
perturbative approach. In this regime unitarity needs to be implemented already at the
level of perturbative interactions and the Glauber-Eikonal model of multiple parton
collisions allows implementing unitarity in the simplest and most natural way. A standard
approximation is to conserve the longitudinal momentum component of the projectile in the
process. The approximation becomes however questionable when dealing with parton
interactions, since parton distributions are singular at small $x$ and a small change of
the parton momentum leads to sizable modifications of the initial parton flux.

In the present note we have studied this particular aspect of the problem, working out
the leading terms in the expansion in multiple parton collisions and taking into account
all kinematical constraints exactly. Our result is that the spectrum of the produced
large $p_t$ partons is appreciably softened after implementing exact kinematics.

The Cronin effect has been recently measured in the large $p_t$ spectrum of $\pi^0$
production in proton-lead collisions, at $y=0$ and $\sqrt s=200$ AGeV. We have worked out
the $\pi^0$ spectrum in the same kinematical conditions and found a rather good agreement
of our model with experimental data. In this case the corrections to the Glauber Eikonal
model, induced by implementing kinematics exactly, are increasingly important in the
region of smaller $p_t$, where, given the rapidity value of the observed $\pi^0$,
longitudinal and transverse momentum components are of comparable magnitude. The
quenching of the spectrum, due to the energy lost by the projectile through multiple
collisions, is a stronger effect at larger rapidity. We have estimated the effect at
$\eta=3.2$, where the preliminary experimental indications seem however to indicate a
stronger quenching effect than expected in the model, which, at relatively large $p_t$,
still gives a value of the Cronin ratio larger that one.
\section{ Acknowledgments}
We are grateful to M.~Tosolini for his valuable computational support. One of us, D.T.,
thanks the Department of Energy's Institute for Nuclear Theory at the University of
Washington for hospitality and the Department of Energy for partial support during the
completion of this work. This work was partially supported by the Italian Ministry of
University and of Scientific and Technological Researches (MIUR) by the Grant COFIN2003.
\section{Appendix}
The four-body parton interaction (one projectile parton against three target partons) is
represented in the model by the convolution of three on-shell elastic parton-parton
scatterings. We call $h$ the four-momentum of the initial state projectile parton, and
$p'$, $p''$,$p'''$, the four-momenta of the three nuclear target partons. All partons are
massless and on shell and have zero transverse momentum components. By introducing the
momentum fractions $x$, $x'$, $x''$, the light-cone components of the four momenta in the
initial state are:

\[
h=\left(\sqrt{s}\, x,0;\bar{0}\,\right)\quad p'=\left(0,\sqrt{s}\, x';\bar{0}\right)\quad
p''=\left(0,\sqrt{s}\, x'';\bar{0}\right)\quad p'''=\left(0,\sqrt{s}\,
x''';\bar{0}\right)\]

The transverse components of the exchanged four-momenta are $\bar{q}_{1}$, $\bar{q}_{2}$
and $\bar{q}_{3}$, while $I$, $J$, $K$ are the projectile's four-momenta after the first,
second and third interaction and $P$, $Q$ and $R$ the four-momenta of the recoiling
target partons. For the transverse components one has:

\[
I_{\perp}=\bar{q}_{1}\quad J_{\perp}=\bar{q}_{1}+\bar{q}_{2}\quad
K_{\perp}=\bar{q}_{1}+\bar{q}_{2}+\bar{q}_{3}\quad P_{\perp}=-\bar{q}_{1}\quad
Q_{\perp}=-\bar{q}_{2}\quad R_{\perp}=-\bar{q}_{3}\]

While the energy conservation and mass-shell conditions, for each two-body scattering
process, are expressed by the following equations:

\begin{equation}
\left\{ \begin{array}{ccc}
I^{+}I^{-} & = & \bar{q}_{1}^{2}\\
P^{+}P^{-} & = & \bar{q}_{1}^{2}\\
\sqrt{s}x & = & I^{+}+P^{+}\\
\sqrt{s}x' & = & I^{-}+P^{-}\end{array}\right.\label{eq:first scattering}\end{equation}

\begin{equation}
\left\{ \begin{array}{ccc}
J^{+}J^{-} & = & \left(\bar{q}_{1}+\bar{q}_{2}\right)^{2}\\
Q^{+}Q^{-} & = & \bar{q}_{2}^{2}\\
I^{+} & = & J^{+}+Q^{+}\\
I^{-}+\sqrt{s}x'' & = & J^{-}+Q^{-}\end{array}\right.\label{eq:second
scattering}\end{equation}

\begin{equation}
\left\{ \begin{array}{ccc}
K^{+}\, K^{-} & = & \left(\bar{q}_{1}+\bar{q}_{2}+\bar{q}_{3}\right)^{2}\\
R^{+}\, R^{-} & = & \bar{q}_{3}^{2}\\
J^{+} & = & K^{+}+R^{+}\\
J^{-}+\sqrt{s}x''' & = & K^{-}+R^{-}\end{array}.\right.\label{eq:trird
scattering}\end{equation} The three systems of equations allow to determine \[
x\,,I^{+},I^{-},J^{+},J^{-},K^{-},P^{+},P^{-},Q^{+},Q^{-},R^{+},R^{-}\]
 as a function of $\bar{q}_{1}$, $\bar{q}_{2}$, $\bar{q}_{3}$,
$K^{+}$, $x'$, $x''$ and $x'''$.

From the system (\ref{eq:trird scattering}) and the mass shell condition
$J^{+}J^{-}=\left(\bar{q}_{1}+\bar{q}_{2}\right)^{2}$, one obtains:

\begin{eqnarray*}
\left(\sqrt{s}x'''\: K^{+}-\left(\bar{q}_{1}+\bar{q}_{2}+\bar{q}_{3}\right)^{2}\right)\,\left(J^{+}\right)^{2}-\left(\bar{q}_{1}+\bar{q}_{2}\right)^{2}\,\left(K^{+}\right)^{2}\\
-\left(\sqrt{s}x'''\,
K^{+}-2\,\left(\bar{q}_{1}+\bar{q}_{2}\right)\cdot\left(\bar{q}_{1}+\bar{q}_{2}+\bar{q}_{3}\right)\right)\,
J^{+}\, K^{+} & = & 0\end{eqnarray*}
 which has real solutions when $\Lambda^{(3)}\geq0$, where

\begin{eqnarray*}
\Lambda^{(3)} & = & \left(\sqrt{s}x'''\, K^{+}\right)^{2}-4\,\left(\bar{q}_{1}+\bar{q}_{2}\right)\cdot\bar{q}_{3}\,\sqrt{s}x'''\, K^{+}\\
 &  & +4\left(\left(\left(\bar{q}_{1}+\bar{q}_{2}\right)\cdot\bar{q}_{3}\right)^{2}-\left(\bar{q}_{1}+\bar{q}_{2}\right)^{2}\,\bar{q}_{3}^{2}\right)\end{eqnarray*}
 The condition $\Lambda^{(3)}\geq0$ can be satisfied only if $\Phi^{(3)}\geq0$,
where:

\begin{equation*}
\Phi^{(3)}=1-2\,\frac{\left(\left(\bar{q}_{1}+\bar{q}_{2}\right)\cdot\bar{q}_{3}+\mid\bar{q}_{1}+\bar{q}_{2}\mid\,\mid\bar{q}_{3}\mid\right)}{\sqrt{s}\,
x'''\, K^{+}}\label{eq:thirdcondition}\end{equation*}
 which gives the kinematical limits for $x'''$ and for $K^{+}$

\begin{equation*}
x'''_{min}\equiv\frac{2\left(\left(\bar{q}_{1}+\bar{q}_{2}\right)\cdot\bar{q}_{3}+\mid\bar{q}_{1}+\bar{q}_{2}\mid\,\mid\bar{q}_{3}\mid\right)}{\sqrt{s}\,
x'''\, K^{+}}\equiv\frac{K_{min}^{+}}{K^{+}}\leq1\label{eq:x''' min}\end{equation*}
 One hence obtains $J$ and $R$:

\begin{eqnarray*}
J^{+} & = & \frac{K^{+}}{2}\cdot\frac{\sqrt{s}\, K^{+}\, x'''-2\,\left(\left(\bar{q}_{1}+\bar{q}_{2}\right)\cdot\left(\bar{q}_{1}+\bar{q}_{2}+\bar{q}_{3}\right)\right)+\sqrt{\Lambda^{(3)}}}{\sqrt{s}\, K^{+}\, x'''-\left(\bar{q}_{1}+\bar{q}_{2}+\bar{q}_{3}\right)^{2}}\\
J^{-} & = & \frac{\sqrt{\Lambda^{(3)}}+2\,\left(\left(\bar{q}_{1}+\bar{q}_{2}\right)\cdot\left(\bar{q}_{1}+\bar{q}_{2}+\bar{q}_{3}\right)\right)-\sqrt{s}\, K^{+}\, x'''}{2\, K^{+}}\\
R^{+} & = & \frac{K^{+}}{2}\cdot\frac{\sqrt{\Lambda^{(3)}}+2\,\left(\bar{q}_{3}\cdot\left(\bar{q}_{1}+\bar{q}_{2}+\bar{q}_{3}\right)\right)-\sqrt{s}\, K^{+}\, x'''}{\sqrt{s}\, K^{+}\, x'''-\left(\bar{q}_{1}+\bar{q}_{2}+\bar{q}_{3}\right)^{2}}\\
R^{-} & = & \frac{\sqrt{\Lambda^{(3)}}+\sqrt{s}\, K^{+}\,
x'''-2\,\left(\bar{q}_{3}\cdot\left(\bar{q}_{1}+\bar{q}_{2}+\bar{q}_{3}\right)\right)}{2\,
K^{+}}\,,\end{eqnarray*}
 while the Mandelstam invariants $\hat{s}_{3}=\left(J+p'''\right)^{2},\,$$\hat{t}_{3}=\left(K-J\right)^{2},\,\hat{u}_{3}=\left(R-J\right)^{2}$
are expressed by:

\begin{eqnarray*}
\hat{s}_{3} & = & \sqrt{s}\, x'''\, J^{+}\nonumber \\
\hat{t}_{3} & =- & \frac{\left(\left(\bar{q}_{1}+\bar{q}_{2}+\bar{q}_{3}\right)\, J^{+}-K^{+}\,\left(\bar{q}_{1}+\bar{q}_{2}\right)\right)^{2}}{K^{+}\, J^{+}}\label{Third-Scattering-Mandelstam-Variables}\\
\hat{u}_{3} & = &
-\frac{\left(J^{+}\bar{q}_{3}+R^{+}\left(\bar{q}_{1}+\bar{q}_{2}\right)\right)^{2}}{J^{+}R^{+}}\nonumber
\end{eqnarray*}

After solving (\ref{eq:second scattering}) with respect to $I$ and $Q$ as a function of
$\bar{q}_{1}$, $\bar{q}_{2}$, $\bar{q}_{3}$, $J^{+}$ and $x''$ and using the mass shell
condition $I^{+}I^{-}=\bar{q}_{1}^{2}$ one obtains:

\begin{eqnarray*}
\left(\sqrt{s}x''\: J^{+}-\left(\bar{q}_{1}+\bar{q}_{2}\right)^{2}\right)\,\left(I^{+}\right)^{2}-\left(\bar{q}_{1}\right)^{2}\,\left(J^{+}\right)^{2}\\
-\left(\sqrt{s}x''\,
J^{+}-2\,\bar{q}_{1}\cdot\left(\bar{q}_{1}+\bar{q}_{2}\right)\right)\, I^{+}\, J^{+} & =
& 0\end{eqnarray*} which has real solutions when $\Lambda^{(2)}\ge0$, where

\begin{eqnarray*}
\Lambda^{(2)} & = & \left(\sqrt{s}x''\,
J^{+}\right)^{2}-4\,\bar{q}_{1}\cdot\bar{q}_{2}\,\sqrt{s}x''\,
J^{+}+4\left(\left(\bar{q}_{1}\cdot\bar{q}_{2}\right)^{2}-\bar{q}_{1}^{2}\,\bar{q}_{2}^{2}\right)\,.\end{eqnarray*}
The condition $\Lambda^{(2)}\ge0$ is satisfied only if $\Phi^{(2)}\geq0$, where

\begin{equation*}
\begin{array}{ccc}
\Phi^{(2)} & = &
1-2\,\frac{\left(\bar{q}_{1}\cdot\bar{q}_{2}+\mid\bar{q}_{1}\mid\,\mid\bar{q}_{2}\mid\right)}{\sqrt{s}\,
x''\, J^{+}}\end{array}\label{eq:second condition}\end{equation*} which gives the
kinematical limits for $x''$ and for $K^{+}$

\begin{equation*}
x''_{min}\equiv\frac{2\left(\bar{q}_{1}\cdot\bar{q}_{2}+\mid\bar{q}_{1}\mid\,\mid\bar{q}_{2}\mid\right)}{\sqrt{s}\,
x''\, J^{+}}\equiv\frac{J_{min}^{+}}{J^{+}}\leq1\label{eq:x'' min}\end{equation*} One
hence obtains I and $Q$:

\begin{eqnarray*}
I^{+} & = & \frac{J^{+}}{2}\cdot\frac{\sqrt{s}\, J^{+}\, x''-2\,\left(\bar{q}_{1}\cdot\left(\bar{q}_{1}+\bar{q}_{2}\right)\right)+\sqrt{\Lambda^{(2)}}}{\sqrt{s}\, J^{+}\, x''-\left(\bar{q}_{1}+\bar{q}_{2}\right)^{2}}\\
I^{-} & = & \frac{\sqrt{\Lambda^{(2)}}+2\,\left(\bar{q}_{1}\cdot\left(\bar{q}_{1}+\bar{q}_{2}\right)\right)-\sqrt{s}\, J^{+}\, x''}{2\, J^{+}}\\
Q^{+} & = & \frac{J^{+}}{2}\cdot\frac{\sqrt{\Lambda^{(2)}}+2\,\left(\bar{q}_{2}\cdot\left(\bar{q}_{1}+\bar{q}_{2}\right)\right)-\sqrt{s}\, J^{+}\, x''}{\sqrt{s}\, J^{+}\, x''-\left(\bar{q}_{1}+\bar{q}_{2}\right)^{2}}\\
Q^{-} & = & \frac{\sqrt{\Lambda^{(2)}}+\sqrt{s}\, J^{+}\,
x''-2\,\left(\bar{q}_{2}\cdot\left(\bar{q}_{1}+\bar{q}_{2}\right)\right)}{2\,
J^{+}}\,,\end{eqnarray*}
 while the Mandelstam invariants $\hat{s}_{2}=\left(I^{+}+p''\right)^{2},\;\hat{t}_{2}=\left(J-I\right)^{2},\:\hat{u}_{2}=\left(I-Q\right)^{2}$are
given by:

\begin{eqnarray*}
\hat{s}_{2} & = & \sqrt{s}\, x''\, I^{+}\nonumber \\
\hat{t}_{2} & = & -\frac{\left(\left(\bar{q}_{1}+\bar{q}_{2}\right)\, I^{+}-J^{+}\bar{q}_{1}\right)^{2}}{I^{+}\, J^{+}}\label{Second-Scattering-Mandelstam-Variables}\\
\hat{u}_{2} & = & \frac{\left(I^{+}\bar{q}_{2}+Q^{+}\bar{q}_{1}\right)^{2}}{I^{+}\,
Q^{+}}\nonumber \end{eqnarray*}

By solving (\ref{eq:first scattering}) with respect to $x$ and $P$ as a function of $x'$,
$I$ and $\bar{q}_{1}$one obtains:

\begin{align*}
x & =\frac{x'\, I^{+}}{\sqrt{s}\, x'-I^{-}}\label{eq:incomingfraction}\\
P^{+} & =\frac{\bar{q}_{1}^{2}}{\sqrt{s}\, x'-I^{-}}\nonumber \\
P^{-} & =\sqrt{s}\, x'-I^{-}\nonumber \end{align*}
 The lower limit of $x'$ is derived when requiring $x$ to be smaller
than 1:

\begin{equation*}
x'\geq x'_{min}\equiv\frac{I^{-}}{\sqrt{s}-I^{+}}\end{equation*}
 so we introduce

\begin{equation*}
\Phi^{(1)}=1-\frac{I^{-}}{x'\,\left(\sqrt{s}-I^{+}\right)}\label{eq: first
condition}\end{equation*} while the expression of the Mandelstam invariants
$\hat{s}_{1}=\left(h+p\right)^{2},\,$ $\hat{t}_{1}=\left(I-h\right)^{2}$ and
$\hat{u}_{1}=\left(P-h\right)^{2}$is:

\begin{eqnarray*}
\hat{s}_{1} & = & s\, x\, x'\nonumber \\
\hat{t}_{1} & = & -\sqrt{s}\, x'\, I^{+}\label{First-Scattering-Mandelstam-variables}\\
\hat{u}_{1} & = & -\sqrt{s}\, x\, P^{-}\nonumber \end{eqnarray*}

\vspace{.5cm}
\begin{center}
\begin{figure}
\includegraphics
[ scale=0.6]{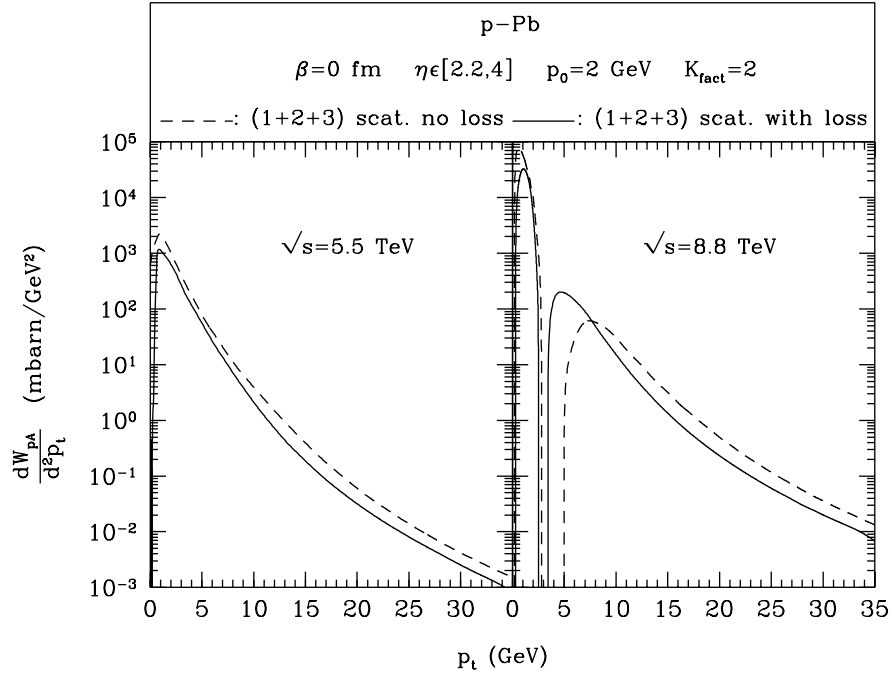}
\caption{{\small Transverse momentum spectrum of partons produced in $p+Pb$ collisions at
$\sqrt{s}=5.5\, ,8.8\, ATeV\,$and $\eta\in[2.4,\,4]$, using  $p_{0}=2\, GeV,\,$ and
$k_{factor}=2\,$. The solid lines are the result of implementing exact kinematics in the
calculation, the dashed lines refer to the case of approximate kinematics.}}
\end{figure}

\begin{figure}
\includegraphics[scale=0.6]{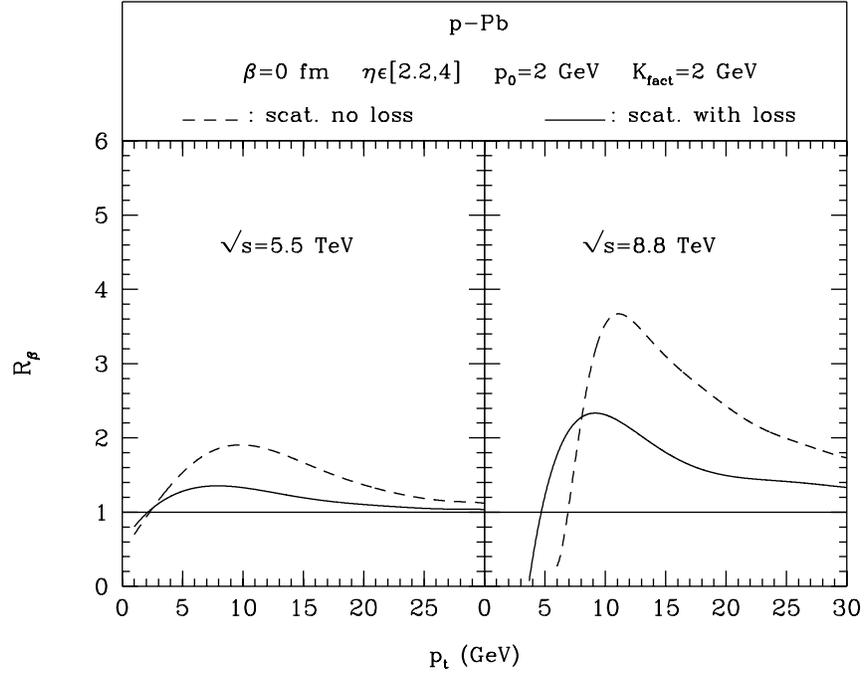}
\caption{Cronin Ratio in the case of exact kinematics (solid line) and in the case of
approximate kinematics (dashed line) with the two different choices of energy scale.}
\end{figure}

\begin{figure}
\includegraphics[
  scale=0.35]{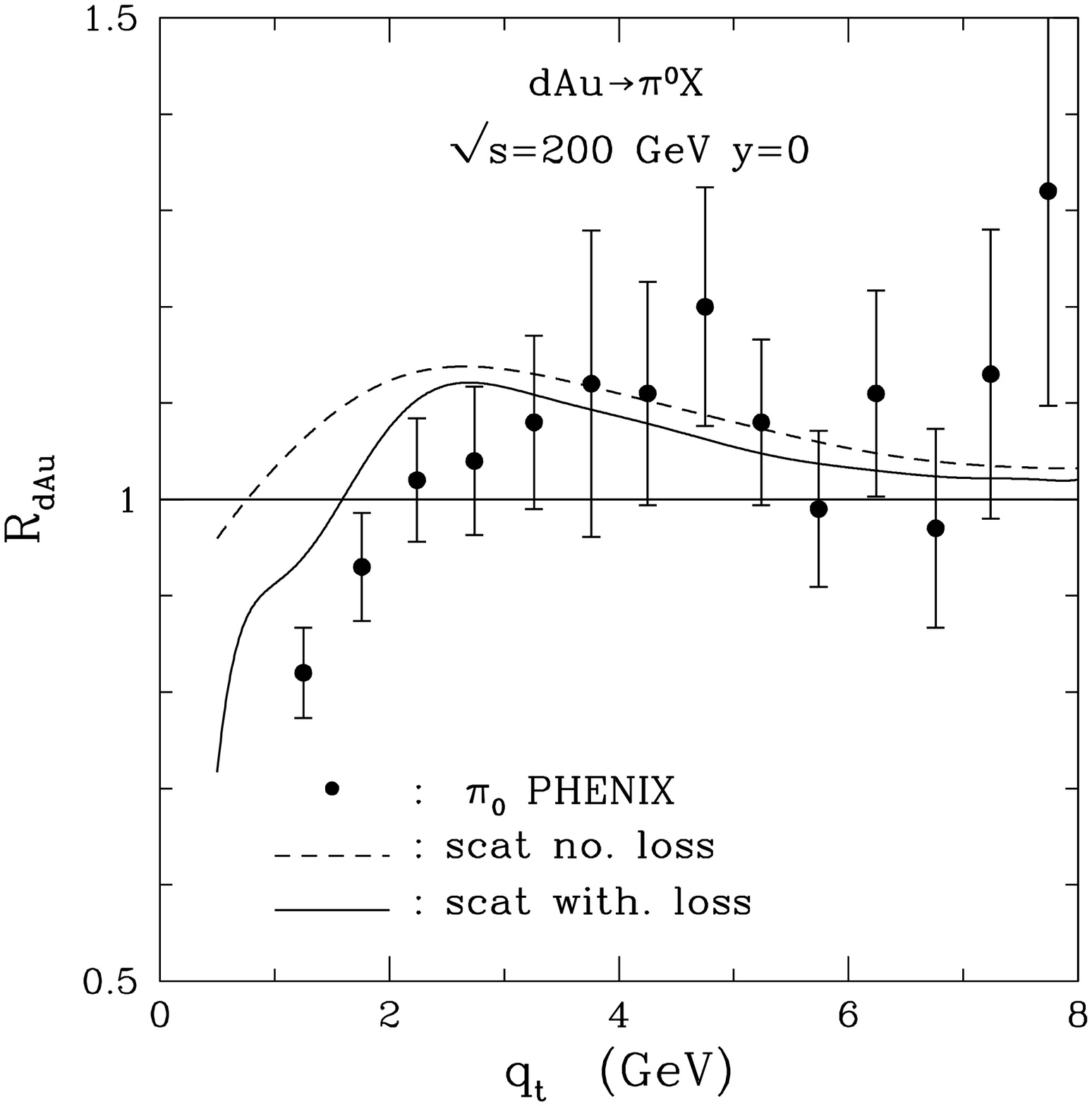}
  \includegraphics[
  scale=0.35]{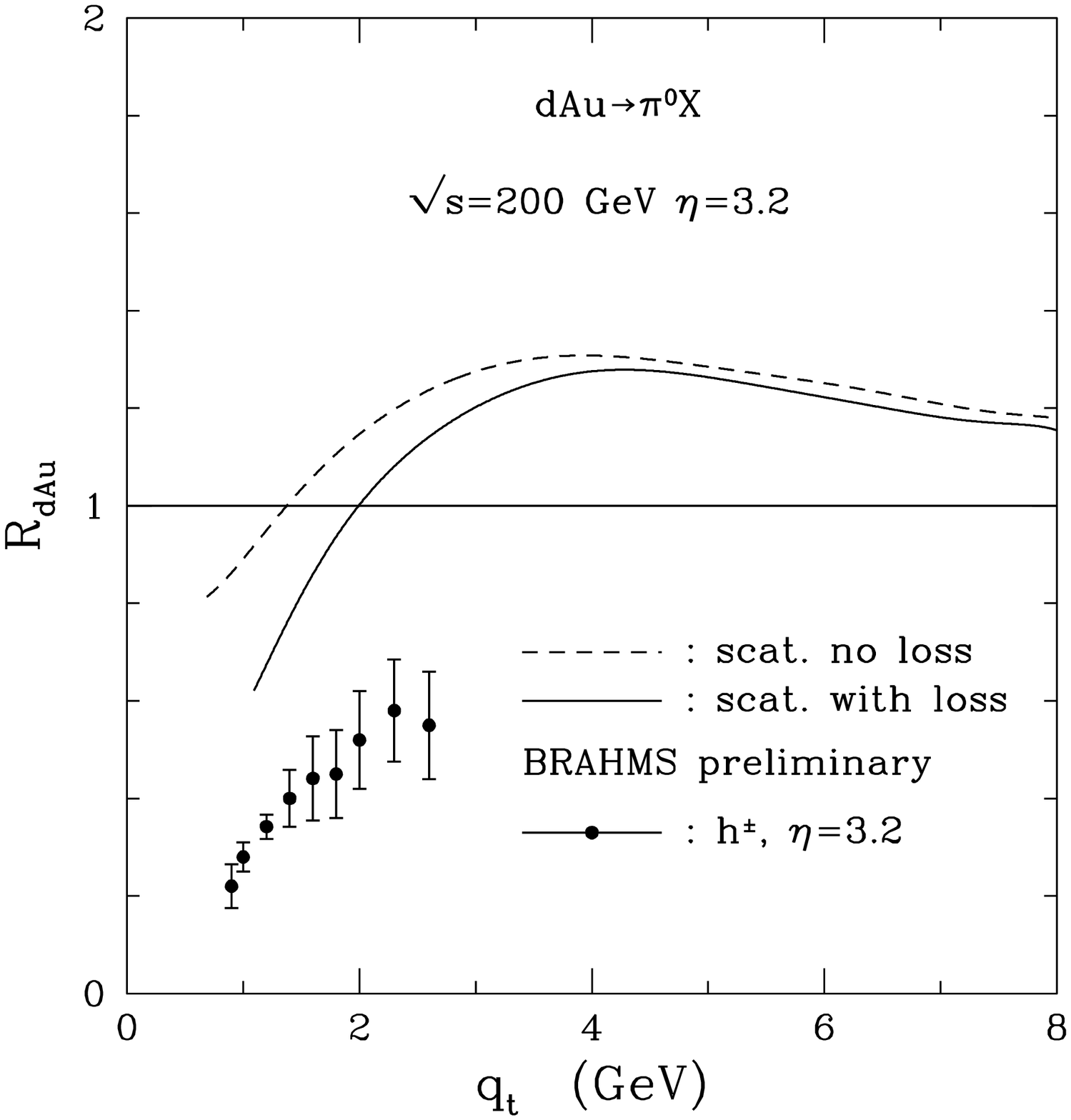}
\caption{Cronin ratio in $d+Au\,\rightarrow\pi^{0}\, X$ collisions at $\sqrt{s}=200\,
GeV\,$. The solid line refers to the case of exact kinematics and the dashed line to the
case of approximate kinematics. Left and right panels refer respectively to $y=0$ and $\eta=3.2$. The data are form ref. \cite{Adler:2003ii} and
\cite{Brahms}.}
\end{figure}

\begin{figure}
\includegraphics[scale=0.6]{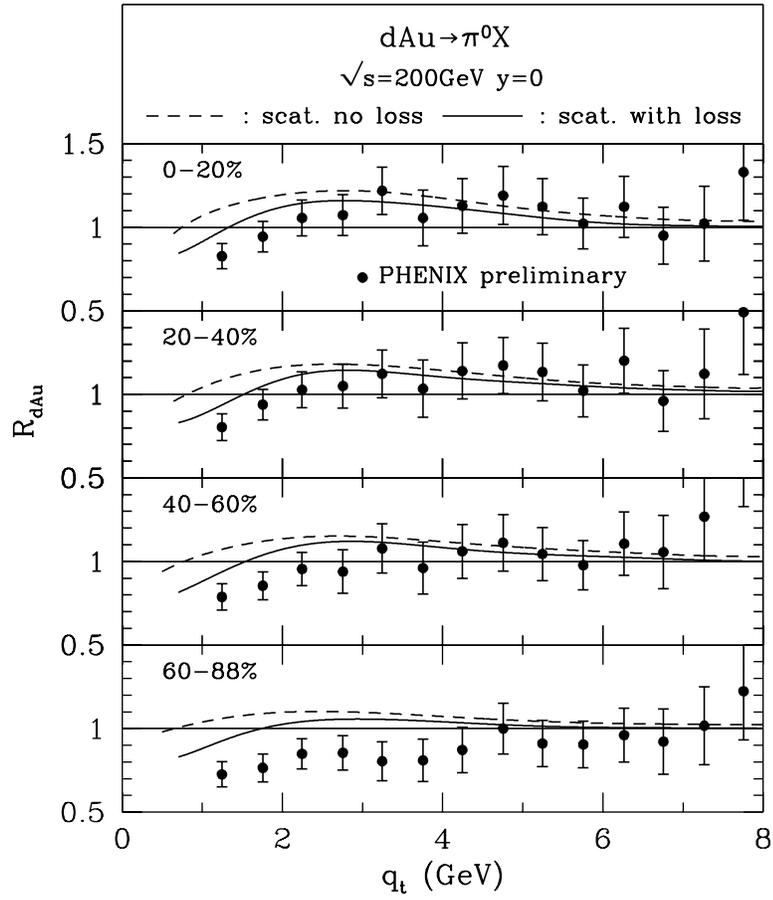}
\caption{Centrality dependence of Cronin ratio with and without energy loss implementation. Comparison with
experimental data presented by PHENIX.}
\end{figure}
\end{center}

\end{document}